\pdfoutput=1
\documentclass[a4paper,twocolumn,epjc3]{svjour3e}
\smartqed 
\makeatletter \let\cl@chapter\relax \makeatother

\usepackage{mathptmx}  
\usepackage[latin1]{inputenc}
\usepackage[T1]{fontenc}
\usepackage[english]{babel}
\usepackage{amsmath,amssymb}
\usepackage[usenames]{color}
\usepackage{graphicx,subfig,afterpage}
\usepackage[numbers,sort&compress]{natbib}
\usepackage[normalem]{ulem}

\usepackage[pdftex]{hyperref}
\hypersetup{colorlinks=true,linkcolor=black,citecolor=black,filecolor=black,
      pdfauthor={M. Cadoni, E. Franzin and S. Mignemi},
      pdftitle={Inflation as de Sitter instability}}
\usepackage[capitalise]{cleveref}
\crefformat{plural}{#2Eqs.~(#1)#3} 
\crefname{section}{Sect.}{Sects.}

\mathchardef\ordinarycolon\mathcode`\:
\mathcode`\:=\string"8000
\begingroup \catcode`\:=\active
 \gdef:{\mathrel{\mathop\ordinarycolon}}
\endgroup

\newcommand*{\refcite}{Ref.~\cite}
\newcommand*{\refscite}{Refs.~\cite}

\renewcommand{\ge}{\geqslant}
\renewcommand{\le}{\leqslant}
\renewcommand{\b}{\beta}
\newcommand{\g}{\gamma}
\renewcommand{\L}{\Lambda}
\renewcommand{\l}{\lambda}
\newcommand{\p}{\partial}

\newcommand{\ep}{\epsilon}

\DeclareMathOperator\Cosh{Cosh}

\makeatletter\g@addto@macro\bfseries{\boldmath}\makeatother%

\newcommand{\0}{\nonumber}
\newcommand{\ie}{\emph{i.e.}}
\newcommand{\eg}{\emph{e.g.}}
\def\be#1\ee{\begin{align}#1\end{align}}
\def\bsube[#1]#2\esube{\begin{subequations}\label[plural]{#1}\begin{align}#2\end{align}\end{subequations}}

\begin{document}

\title{Inflation as de Sitter instability}

\author{Mariano Cadoni\thanksref{addr1,addr2,e1} \and Edgardo Franzin\thanksref{addr1,addr2,e2}\and and Salvatore Mignemi\thanksref{addr3,addr2,e3}}

\thankstext{e1}{\href{mailto:mariano.cadoni@ca.infn.it}{\color{blue}mariano.cadoni@ca.infn.it}}
\thankstext{e2}{\href{mailto:edgardo.franzin@ca.infn.it}{\color{blue}edgardo.franzin@ca.infn.it}}
\thankstext{e3}{\href{mailto:smignemi@unica.it}{\color{blue}smignemi@unica.it}}

\authorrunning{M.~Cadoni, E.~Franzin and S.~Mignemi}

\institute{Dipartimento di Fisica, Universit\`a di Cagliari, Cittadella Universitaria, 09042 Monserrato, Italy\label{addr1}
      \and
      INFN, Sezione di Cagliari, Cittadella Universitaria, 09042 Monserrato, Italy\label{addr2}
      \and
      Dipartimento di Matematica e Informatica, Universit\`a di Cagliari, viale Merello 92, 09123, Cagliari, Italy\label{addr3}}

\date{}

\maketitle

\emergencystretch=1em

\begin{abstract}
We consider cosmological inflation generated by a scalar field slowly rolling off from a de Sitter
maximum of its potential. The models belong to the class of hilltop models and represent
the most general model of this kind in which the scalar potential can be
written as the sum of two exponentials. The minimally coupled Einstein-scalar gravity theory obtained in this
way is the cosmological version of a two-scale generalization of known holographic models,
allowing for solitonic solutions interpolating between an AdS spacetime in the infrared and scaling solutions in
the ultraviolet. We then investigate cosmological inflation in the slow-roll approximation. Our model reproduces
correctly, for a wide range of its parameters, the most recent experimental data for the power spectrum of primordial perturbations.
Moreover, it predicts inflation at energy scales of four to five orders of magnitude below the Planck scale.
At the onset of inflation, the mass of the tachyonic excitation, \ie\ of the inflaton, turns out to be seven to eight orders of
magnitude smaller than the Planck mass.
\end{abstract}

\section{Introduction\label{sect:intro}}

Nowadays, inflationary cosmology~\cite{Starobinsky:1980te,Guth:1980zm,Sato:1980yn,Linde:1981mu,Albrecht:1982wi} represents the easiest way to solve
the problems of the standard Friedmann-Robertson-Walker (FRW) cosmology, such as the horizon and flatness problems --- for a review, see \eg~\refcite{Linde:2007fr}.

The simplest way to generate inflation is to minimally couple Einstein gravity to a scalar field (the inflaton)
with a self-interaction potential. There exist a plethora of single field inflationary models that can be classified according
to the features of the potential~\cite{Dodelson:1997hr}.
Other alternatives include more scalar fields, as in the curvaton mechanism~\cite{Enqvist:2001zp,Lyth:2001nq,Moroi:2001ct}.

Nevertheless, the most recent data of the Planck satellite exclude non-Gaussian perturbations and
give a striking experimental confirmation of the simplest single-field inflationary
scenario~\cite{Ade:2013zuv,Planck:2013jfk,Ade:2013ydc,Ade:2015tva,Ade:2015ava,Ade:2015lrj},
and in particular the Starobinsky model~\cite{Starobinsky:1980te,Starobinsky:1983zz,Whitt:1984pd},
or more in general, the so-called cosmological attractors~\cite{Downes:2012xb,Kallosh:2013tua,Kallosh:2013yoa,Linde:2014nna},
characterized by a ``red'' power spectrum for primordial perturbations and a small tensor/ scalar amplitude ratio.

The accuracy of the observational data concerning the power spectrum of primordial quantum fluctuations represents
an efficient guide to select inflation models. But, despite the recent remarkable improvements, the important questions
about the microscopic origin of the inflaton and about the physics before inflation are still unanswered.
This lack of knowledge does not allow one to single out a unique inflationary model, \ie\ a specific form of the potential.
In fact, although the Planck data can be used to strongly constrain the inflationary model, mainly through the values of the spectral index~$n_s$
and the tensor/scalar amplitude ratio~$r$, they are not sufficient to select a unique model.

In view of this situation, it is natural to look for hints coming from somewhere else in gravitational physics,
for instance supergravity and string theory~\cite{Burgess:2013sla,Kallosh:2013hoa,Fre:2013vza,Binetruy:2014zya}.

In recent times, minimally coupled Einstein-scalar gravity has been intensively investigated
for holographic applications~\cite{Cadoni:2011nq,Cadoni:2011yj,Cadoni:2012uf,Cadoni:2012ea,Cadoni:2013hna,Roychowdhury:2015cta}.
A class of Einstein-scalar gravity models of particular interest are those allowing for solitonic solutions
interpolating between anti de Sitter (AdS) vacua and domain wall~(DW) solutions with scale-covariant symmetries.
The holographically dual QFT has scaling symmetries, which have a nice interpretation in terms of features of
phase transitions in condensed matter systems (hyperscaling violation).
These solitonic solutions are naturally related to cosmological solutions by the so-called DW/cosmology duality,
a sort of analytic continuation, which maps the soliton in a FRW solution~\cite{Skenderis:2006jq,Skenderis:2007sm,Cadoni:2013gza}.

The cosmological duals of solitons which interpolate between an AdS spacetime at large distances of the bulk theory
(the ultraviolet of the dual QFT) and a scale-covariant geometry at small distances in the bulk theory (the infrared of the dual QFT)
are natural candidates for describing dark energy~\cite{Cadoni:2013hna}.
On the other hand, the cosmological duals of solitons interpolating between AdS in the infrared and scale-covariant geometries in the
ultraviolet~\cite{Cadoni:2011yj,Cadoni:2012uf} may be relevant for describing inflation.
It has been shown that the cosmological solutions of this class of models generate inflation as the scalar field
rolls down from a de Sitter (dS) spacetime~\cite{Mignemi:2014kea}. In this context, inflation can be described as an
instability of the dS spacetime rolling down to a scaling solution.
Such models are known as hilltop models~\cite{Boubekeur:2005zm,Kohri:2007gq}
and inflation is generated by a scalar field rolling off from a local maximum to the potential.
In such a scenario, since inflation starts from a local maximum, the slow-roll conditions can be
satisfied more easily. On the experimental side, hilltop models are a subset of the small-field models,
which are characterized by a potential with negative curvature.
This behaviour of the potential is typical of spontaneous symmetry breaking and phase transitions, \eg\ quartic potentials,
natural inflation models~\cite{Freese:1990rb} and Coleman-Weinberg potentials~\cite{Linde:1982zj}.
Although hilltop models have been widely used to generate cosmological inflation, in most of them the potential is constructed using powers of the the inflaton field. To our knowledge, little attention has been given to hilltop models in which
the potential is built as a combination of two exponentials. In this paper we discus the most general, holographically motivated, hilltop model, for which 
the potential can be written as the sum of two exponentials. We will show that although near the maximum our model has the well-known behaviour of hilltop models with a parabolic potential, at late times it gives predictions of the spectral parameters of the CMB radiation, which are specific for a two-exponential potential.

The structure of the paper is as follows.
In \cref{sect:s2} we generalize the model proposed in \refcite{Mignemi:2014kea} and we construct the most general potential given by
the sum of two exponentials. We show that the minimally coupled
Einstein-scalar gravity theory constructed in this way is the cosmological version of a two-scale generalization of the holographic
models of \refscite{Cadoni:2011yj,Cadoni:2012uf}. In \cref{sect:s3} we discuss the cosmological solution of our model.
Inflation and the spectral parameters of the power spectrum of primordial perturbations are discussed in \cref{sect:slowroll}
using the slow-roll approximation. In \cref{sect:s5} we compare the theoretical predictions of our model with observations.
Finally, in \cref{sect:s6} we state our conclusions and in \ref{sect:a} we briefly repeat our calculations for a model in
which the potential has a constant additive term.

\section{The model\label{sect:s2}}

The simplest way to fuel inflation into a cosmological scenario is to couple, minimally, Einstein gravity
to a scalar field~$\phi$ with an appropriate self-interaction potential $V(\phi)$:
\be\label{action}
A=\int{}d^4x\sqrt{-g}\left(\frac{m_P^2}{16\pi}\,R -\frac{1}{2}(\p\phi)^2-V (\phi)\right).
\ee%

In this paper we focus on inflation generated by a scalar field rolling off from a maximum of $V$.
This class of models is very natural from a physical point of view because inflation can be thought of just as
an instability of the dS spacetime, generated by a scalar perturbation.

Our first goal is to construct the general form of the potential belonging to this class.
Without loss of generality we can assume that the maximum of the potential occurs at $\phi=0$, so that the basic
necessary conditions to be imposed on the potential read
\be\label[plural]{cond}
V (0)>0,\quad V' (0)=0,\quad V'' (0)<0.
\ee%
Obviously, the previous conditions are very loose and do not select any specific form of $V(\phi)$.
We further constrain the form of the potential by requiring it to be a linear combination of two
exponentials. This is a rather strong assumption, but is supported by several arguments.
Exponential potentials for scalar field appear quite generically in a variety of situations:
compactification of extra dimensions, $f(R)$ gravity theories (which on-shell are equivalent
to Einstein-scalar gravity) and low-energy effective string theory. The double exponential potential 
appears in the context of dimensional reduction of gravity with non-trivial four-form flux on a maximally
symmetric internal space --- see \eg\ \refcite{Jarv:2004uk} and references therein.
Moreover, exponential potentials have been shown to be the source of brane solutions of Einstein-scalar gravity
called DWs~\cite{Cadoni:2011nq,Cadoni:2011yj,Cadoni:2012uf,Cadoni:2012ea},
which can be analytically continued into FRW cosmological solutions~\cite{Cadoni:2013gza,Mignemi:2014kea}.

We are therefore led to consider the following general form of the inflation potential\footnote{One could also consider a potential with an added constant term.
This case will be discussed in \ref{sect:a}.}
\be\label{pot}
V (\phi) =\L^2\left(a_1e^{b_1\mu\phi}+a_2 e^{b_2\mu\phi}\right),
\ee%
where $\L$ and $\mu$ are some length scales, whose physical meaning will be clarified in short, and $a_{1,2},\,b_{1,2}$ are
some dimensionless constants characterizing the model. They are constrained by \cref{cond}, giving
\be\label[plural]{cond1}
a_1+a_2>0,\quad a_1b_1=- a_2b_2,\quad a_1b_1^2+a_2b_2^2 <0.
\ee%
Modulo trivial symmetries interchanging the two exponentials in the potential,
the most general solution of the previous equations is $a_1>0$, $a_2<0$, $b_2>0$, $b_1>0$ and $a_2/a_1=-\b^2$,
where we have defined a new dimensionless parameter $\b^2\equiv b_1/b_2<1$.
The parameter rescaling $\L^2\to 2\L^2/(3 a_2\g)$, $\mu\to\sqrt{3/(b_1b_2)}\,\mu$ brings the potential in the form
\be\label{pot1}
V (\phi) =\frac{2\L^2}{3\g}\left(e^{\sqrt{3}\b\mu\phi}-\b^2 e^{\sqrt{3}\mu\phi/\b}\right),
\ee%
where $\g\equiv1-\b^2$.
The potential \eqref{pot1} is a two-scale generalization of the model proposed in \refscite{Cadoni:2011yj,Cadoni:2012uf},
to which it reduces for the particular value of the parameter $\mu=4\sqrt{\pi}\,l_P$. The cosmology of the latter model has
been investigated in \refcite{Mignemi:2014kea}.\footnote{Notice that our notation differs from that of \refcite{Mignemi:2014kea}
for the units used and for a rescaling of the parameter $\mu$ by a factor of~$2$.}
We will see in the next section that for
generic values of the parameter $\mu\neq 4\sqrt{\pi}\,l_P$ the cosmological equations resulting from
the model \eqref{pot1} do not give rise to an exactly integrable system.

The potential \eqref{pot1} is invariant both under the transformation $\b\to 1/\b$, which corresponds to interchanging the two exponentials
in the potential \eqref{pot1} and under the transformation $\b\to-\b,\,\phi\to-\phi$. These symmetries allow us to limit our consideration to $0<\b<1$.
The two limiting cases $\b=0,1$ correspond, respectively, to a pure exponential 
and to a potential behaving at leading order as
\be
V=(2\Lambda^2/3)\left(1-\sqrt{3}\mu \phi\right) e^{\sqrt{3}\mu\phi}.
\ee
The potential $V(\phi)$ has a maximum at $\phi=0$ corresponding to an
unstable dS solution with $V(0)=(2/3)\L^2$ and a corresponding tachyonic excitation, the inflaton.

\begin{figure}[h]
\centering
\includegraphics[width=0.35\textwidth]{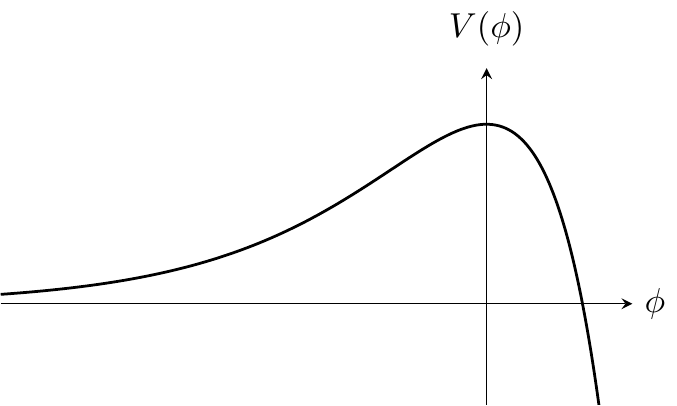}
\caption{Plot of the potential \eqref{pot1} for the following values of the parameters in Planck units:
 $\Lambda=2,\,\beta=3/4,\,\mu=\sqrt{3}/3$.}
\label{fig:potV}
\end{figure}

The potential $V(\phi)$ is depicted in~\cref{fig:potV} for selected values of the parameters $\Lambda$, $\beta$ and $\mu$.

One can therefore use this model to describe inflation as generated by an unstable dS solution.
Inflation starts as a quantum fluctuation of the dS solution and is initially driven
by a tachyonic excitation of the dS spacetime and proceeds as the scalar field rolls off
from the maximum of the potential.

\subsection{Physical scales}
Besides the Planck length $l_P\,=\,1/m_P$, the model is para\-metrized by the two length scales $\L^{-1/2}$ and $\mu$ and by the dimensionless parameter
$\b$. The presence of two length scales is a characteristic feature of small-field models of inflation.
In the present context the two scales have a simple interpretation in terms of geometric properties of the function $V(\phi)$.
They give, respectively, the height and the curvature of the $\phi=0$ maximum of the function $V(\phi)$.
Correspondingly, $\L^{-1/2}$ and $\mu$ determine the two physical scales relevant for inflation:
the vacuum energy $E_V$ at the beginning of inflation and the inflaton mass squared $M_I^2$. We have
\be%
M^2_I &= V'' (0)=-2\L^2\mu^2= -\frac{32\pi}{3}\frac{\l^4}{h^2}m_P^2,\label{pm-mass}\\
E_V &= [V (0)]^{1/4}= (2/3)^{1/4}\l\,m_P,\label{pm-energy}
\ee%
where we have introduced the two dimensionless parameters $h^{-1}$ and $\l$,
\be%
h= 4\,\sqrt{\frac{\pi}{3}}\left(\frac{l_P}{\mu}\right),\quad\l=\frac{\L^{1/2}}{m_P},
\ee%
representing the measures of $\mu$ and $\L^{1/2}$ in Planck units.

Conversely, $\b$ is a purely dimensionless parameter and plays a role which is drastically different from $\l$ and $h$.
It is not linked to any physical scale of the model
but quantifies the deviation of the potential from a pure exponential behaviour attained for $\b$ near 0.

In the following we use instead of the negative quantity $M^2_I$, the inflaton mass defined as $m_I=\sqrt{-M_I^2}$.

\section{Cosmological solutions\label{sect:s3}}

The cosmology of our model can be investigated using the usual FRW parametrization of the metric
\be\label{para1}
ds^2=- dt^2+ a(t)^2 dS_{(3)}^2,
\ee%
The dS spacetime with constant inflaton is an exact solution of cosmological equations.
The dS solution has the usual exponential form, with the scale factor given by
\be\label{ds}
a=e^{8 l_P \sqrt{\pi}\L t/3}.
\ee%
This solution describes a scalar field sitting forever at the maximum
of the potential, generating an exact exponential expansion of the universe, \ie\ never ending inflation.

The most interesting cosmological solutions are those describing inflation lasting for a finite amount of time.
In this case the scalar rolls off from the maximum of $V$, generating a quasi-exponential expansion of the universe
as long as the potential energy of the scalar
dominates the kinetic one. This kind of solutions would be the cosmological counterpart of the solitonic solutions interpolating between
an AdS spacetime in the infrared and a DW in the ultraviolet~\cite{Cadoni:2011yj,Cadoni:2012uf}.

Searching for these solutions, following \refcite{Mignemi:2014kea}, one can try to find exact cosmological solutions by using
a different parametrization for the time variable and 
linear combinations of the fields in such way that the equations for the scalar field and for the scale factor decouple.
However, one can easily realize that the decoupling works only for the particular value of the parameter
$\mu=4\sqrt{\pi}\,l_P$ (corresponding to $h=1/\sqrt{3}$).
For this value of $\mu$ the Einstein-scalar gravity models give rise to exactly integrable models
both in the case of static (brane)~\cite{Cadoni:2011yj,Cadoni:2012uf} and cosmological solutions~\cite{Mignemi:2014kea}.
In the static case we have solitonic solutions interpolating between an AdS spacetime in the infrared and
a DW in the ultraviolet~\cite{Cadoni:2011yj,Cadoni:2012uf,Cadoni:2012ea}. Analogously, in the cosmological case we have
exact solutions which can be used to model inflation~\cite{Mignemi:2014kea}.

For generic values of the parameter $\mu$ the Einstein-scalar system does not decouple,
is not exactly integrable and a cosmological solution cannot be found in analytic form.

Approximate solutions of the field equations can be found for some limiting cases.
Of particular interest is the case of small $\b$, for which the potential \eqref{pot1} behaves exponentially,
\be\label{pot2}
V (\phi) \sim-\frac{2\b^2\L^2}{3\g}e^{\sqrt{3}\mu\phi/\b},
\ee%
the system can be solved analytically and we have scaling (power-law) solutions, which are obtained from scale-covariant (DW)
solutions~\cite{Cadoni:2011nq} using the transformation \mbox{$t\to ir$}, $r\to it$. In the gauge \eqref{para1} this scaling solution has the form
\be\label{ss}
a\propto t^{h^2\b^2},\quad e^{2\phi}\propto t^{-\frac{h\b}{l_P \sqrt{\pi}}}.
\ee%

\section{Inflation and slow-roll approximation\label{sect:slowroll}}

Lacking exact solutions to investigate the cosmology of our model \eqref{pot1}, we work in the slow-roll approximation~\cite{Liddle:2000cg}.
In this regime the potential energy of the scalar field dominates over the kinetic energy and the universe has a quasi-exponential
accelerated expansion as the scalar field slowly rolls off from the maximum of the potential.
Following the usual approach, we introduce the slow-roll parameters $\ep$ and~$\eta$,
\be\label[plural]{sl}
\ep=\frac{m_P^2}{16\pi}\left(\frac{V'}{V}\right)^2,\quad \eta= \frac{m_P^2}{8\pi}\frac{V''}{V}-\ep.
\ee%
Necessary conditions for the slow-roll approximation to be valid are $\ep,|\eta|\ll1$. We have inflation as long as $0\le\ep<1$.
The parameter $\ep$ is zero on the maximum of the potential ($\phi=0$) and 
the solution is exactly dS, whereas inflation ends when $\ep=1$.

The potential \eqref{pot1} is not a monotonic function of the scalar field $\phi$:
it goes to zero as $\phi\to-\infty$, has a maximum at $\phi=0$,
and crosses the axis for $\phi=\phi_*\equiv-\frac{2\b\ln\b}{\sqrt{3}\g\mu}$;
$V\to-\infty$ as $\phi\to\infty$ (see~\cref{fig:potV}).
Since slow-roll inflation cannot occur for a negative inflaton potential,
our model is valid up to $\phi=\phi_*$ and the potential must be modified for values of $\phi$ greater than $\phi_*$.
We have then two alternative branches that we can use to generate inflation,
\ie\ I\@: $0\le\phi\le\phi_*$ and II\@:~$-\infty<\phi\le0$.
In the following, we investigate the first branch and in \cref{sect:sb} we briefly discuss branch II and show that it cannot be compatible with observations.

Let us now introduce the variable
\be\label{nv}
Y= e^{\sqrt{3}\g\mu\phi/\b}.
\ee%
In this parametrization the branch under consideration corresponds to $1\le Y\le Y_*\equiv1/\b^2$.

As a function of $Y$, the slow-roll parameters $\ep$ and $\eta$ take the form
\be\label[plural]{sr}
\ep=\frac{\b^2}{h^2}\left(\frac{1-Y}{1-\b^2 Y}\right)^2,\quad
\eta=\frac{2}{h^2}\frac{\b^2-Y}{1-\b^2 Y}-\ep.
\ee%
The slow-roll parameter $\ep$ is zero for $Y=1$, whereas \mbox{$0<\ep<1$} for $1<Y<Y_0$,
where
\be\label{ooo}
Y_0=\frac{\b+h}{\b+\b^2 h}.
\ee%
For $Y=Y_0$ we have $\ep=1$ and the universe exits
inflation. One can easily check that $Y_0<1/\b^2$,
so that during inflation we always have $1\le Y\le1/\b^2$ and we can easily satisfy the first slow-roll
condition $\ep\ll1$.
On the other hand, the parameter~$\eta$, which gives a measure of the
curvature of the potential, is not small, but we have $\eta=\mathcal{O}(h^{-2})$.
It follows that the simplest way to satisfy the second slow-roll
condition, $|\eta|\ll1$, is to choose
\be\label{jj}
h\gtrsim10,
\ee%
in this way we can have $\eta\approx10^{-2}$ as well as $\ep\approx10^{-2}$. As already noted, the model discussed
in \refcite{Mignemi:2014kea} does not satisfy~\cref{jj} because it is characterized by $h=1/\sqrt{3}$.

In the slow-roll regime, the universe expands quasi-expo\-nentially and the number of $e$-folds $N=-\log{}a$, which
determines the duration of inflation, is determined by
\be\label{ll}
N=-\int dt H=\frac{8\pi}{m_P^2}\int_{\phi_\text{e}}^{\phi_\text{b}}d\phi\,\frac{V}{V'},
\ee%
where $\phi_{\text{e},\text{b}}$ are, respectively, the inflaton-field values at
the end and beginning of inflation and $H=\dot{a}/a$ is the Hubble parameter.

Using the definition \eqref{nv} and the expression $Y_{0}$ for $Y$ at
the end of inflation, \cref{ll} gives the function $Y(N)$ in implicit form,
\be\label{klh}
\frac{Y^{1/\g}}{Y-1}= e^{2N/h^2} A,\quad
A := \frac{\b}{\g}\left(\b+\frac{1}{h}\right)\left(\frac{\b+h}{\b+\b^2 h}\right)^{1/\g}.
\ee%
In the case of the dS solution \eqref{ds} the scalar field remains constant (the inflaton sits on the top of the potential),
and we have $N=\infty$ (eternal inflation). Obviously this configuration is highly unstable. A small perturbation of the scalar
field starts the slow roll of the inflaton along the slope and a finite value of $N$ is generated. If this fluctuation is small
enough we can solve approximately \cref{klh} for $Y$ near $Y=1$. We get at leading order,
\be\label{qrt}
Y= 1+ A^{-1} e^{-2N/h^2}.
\ee%

One can easily check that $0\le A^{-1}\le1$ with $A^{-1}\to0$ for $\b\to1$ and $A^{-1}\to 1$ for $\b\to0$. Moreover, in the range
\mbox{$0\le\b\le1$}, $A^{-1}(\b,h)$ is a monotonically decreasing function of $\b$ which depends very weakly on $h$. It follows immediately
that \cref{qrt} is a good approximation for $\g$ not too close to 0, whenever $e^{-2N/h^2}\ll1$.
When $\g\approx0$ the approximation~\eqref{qrt} holds irrespectively of the value of $N$.

\subsection{Perturbations and spectral parameters\label{sect:s4}}

One of the most striking predictions of inflation concerns the spectrum of tensor and scalar perturbations
in the early universe~\cite{Starobinsky:1979ty,Mukhanov:1981xt,Kodama:1985bj,Mukhanov:1990me,Mukhanov:2013tua}.
During inflation the horizon shrinks and the primordial perturbations, which were causally connected are redshifted to superhorizon scales.
Conversely, in the matter-radiation dominated era the horizon grows, the perturbations fall back in the horizon so that they can act as seeds for
structure formation and anisotropy in the universe. The information as regards these primordial fluctuations is therefore encoded in the an\-isotropies of the CMB\@.

Primordial quantum fluctuations are described in terms of two-point correlation functions for scalar and tensor modes
in Fourier space and the associated power spectrum. In the slow-roll approximation, the power spectrum has a power-law behaviour
and is usually characterized by four parameters: the amplitudes of scalar perturbations $P_R$, the ratio $r$ of the amplitudes
of tensor and scalar perturbations and their spectral indices $n_s$ and $n_T$. These parameters are functions of the number of $e$-folds $N$
and can be expressed in terms of the potential $V$ and the slow-roll parameters \eqref{sr} as follows:
\bsube[param0]%
&P^{1/2}_R (N) = \frac{4\sqrt{24\pi}}{3 m_P^3}\frac{V(\phi(N))^{3/2}}{V'(\phi(N))}\label{pr0},\\
&r (N) =-8n_T(\phi(N)) = 16 \ep(\phi(N))\label{r0},\\
&n_s (N) = 1 -4\ep(\phi(N)) +2\eta(\phi(N)),\label{ns0}
\esube%
where $\phi(N)$ is defined by \cref{ll}.

Using Eqs.~\eqref{nv} and \eqref{sr} we can express the spectral parameters as a function of $Y(N)$:
\bsube[param]%
&P^{1/2}_R (N) =\frac{4 h \l^2}{3\b\sqrt\g}\frac{\left[1-\b^2 Y (N)\right]^{3/2}}{1-Y (N)}\,Y(N)^{\b^2/2\g},\label{pr}\\
&r (N) = \frac{16\b^2}{h^2}\left(\frac{1-Y (N)}{1-\b^2 Y (N)}\right)^2,\label{r}\\
&n_s (N) = 1- \frac{6\b^2}{h^2}\left[\frac{1-Y(N)}{1-\b^2 Y(N)}\right]^2+\frac{4}{h^2}\frac{\b^2-Y (N)}{1-\b^2 Y (N)},\label{ns}
\esube%
where $Y(N)$ is defined, implicitly, by \cref{klh}.

For $e^{-2N/h^2}\ll1$ we can use the approximate expansion for $Y$ given by \cref{qrt} and we get,
at leading order in the $e^{-2N/h^2}$ expansion,
\bsube[param1]
&P^{1/2}_R (N) =\frac{4\g A}{3\b} h \l^2 e^{2N/h^2},\label{pr1}\\
&r (N) =\left(\frac{4\b}{A\g h}\right)^2 e^{-4N/h^2},\label{r1}\\
&n_s (N) =1-\frac{4}{h^2}\left(1+\frac{1+\b^2}{A\g}e^{-2N/h^2}\right).\label{ns1}
\esube%
One important feature of \cref{param1} is the exponential dependence on $N$. This must be compared with the
typical behaviour of the Starobinsky model and more in general of cosmological attractor models, where one
typically obtains $r\propto 1/N^2$ and $n_s-1\propto-1/N$ --- see \eg\ \refcite{Linde:2014nna} and references therein.

It is easy to check that the exponential behaviour of the spectral parameters \eqref{param1} is an universal feature of hilltop models characterized by a quadratic maximum. It is a consequence of the local behaviour of the potential near 
$\phi=0$. In fact \cref{param1} can also be obtained by
considering a potential $V= 2\Lambda^2/3+ M^2_I\phi^2/2$, with $M^2_I$ given by \cref{pm-mass}. This is consistent with the
fact that for $N/h^2$ very large, inflation occurs near to the maximum of the potential, where $V$ can be approximated by the previous form.

Notice that the condition $h\gg1$ alone does not does not guarantee the potential to 
be well approximated by the parabolic one. Since we need at least $h\gtrsim10$, such limit is obtained for $N\gg60$.
For instance for $h=10$ and $N=60$ we have $e^{-2N/h^2}\approx 0.54$. It follows that the approximate expressions~\eqref{param1}
can only be used in a regime of very large $N$, for which we do not have a direct access to observations,
and, therefore, in the following we will be using expressions~\eqref{param}.

\section{Comparison with observation\label{sect:s5}}

In this section we compare the theoretical results of our model for the spectral parameters
$P_R$, $r$ and $n_s$ with the most recent results of observations, in particular the joint analysis of BICEP2/Keck Array and 
Planck data~\cite{Ade:2015tva}.

The spectral parameters are functions of the number of the $e$-folds $N$ and depend on the three dimensionless parameters $\l$, $h$ and $\b$.
Because $\l$ enters only in the normalization of the power spectrum $P_R$, whereas $r$ and $n_s$ depend on $h$ and $\b$ only 
we will use the following strategy: we will first determine using \cref{r,ns} and the experimental results for $r$ and $n_s$,
the allowed range of the parameters $h$ and $\b$ for a given value of $e$-folds~$N$. We will then use \cref{pr} and the experimental results for $P_R$ to
determine the corresponding values of the parameter~$\l$.
Finally we use \cref{pm-mass,pm-energy} to determine the vacuum energy $E_V$ and the inflaton mass $m_I$.

For $r$, $n_s$ and $P_R$ we use the most recent results~\cite{Ade:2015tva}, \ie\ $r<0.05$, $n_s=0.965\pm0.006$ and $P_R^{1/2}\approx10^{-5}$.
Since the perturbations we are observing today with momentum of the order of the horizon radius 
exited the horizon during inflation at $N=[48,60]$, we will consider only values of $N$ in this range.

The calculations have to be performed numerically because the function $Y(N)$ appearing in \cref{param}
is not known, but it is defined implicitly by \cref{klh}.
As we said in the previous section, a possible way to avoid numerical computations is to work in a regime where $e^{-2N/h^2}\ll1$
and then \cref{param1} hold. But unfortunately, these expressions are valid in the large $N$ regime, not accessible to observations.

The results of our numerical computations are shown in the two sets of density plots in \cref{fig:rns,fig:EVmI}.
Once we have chosen the value of $N$, the coloured region in such plots represent the range of values of $\b$ and $h$ for which we have values
of~$r$ and~$n_s$ compatible with the experimental measurements.

Note that the allowed region of parameters $(\b,h)$ is quite independent from $N$, at least for $N$ in the range $[48,60]$. 

\subsection{Spectral parameters}

In \cref{fig:rns} we show the numerical results obtained from Eqs. \eqref{r} and \eqref{ns}. We plot the tensor/scalar ratio~$r$
(left) and the spectral index $n_s$ (right) as functions of $\b$ and $h$ for four selected values of $N=48,52,56,60$.
The corresponding values of $r$ and~$n_s$ are given in terms of the colour scale shown on the right of every plot.

In general, higher values of $n_s$ correspond to higher values of $h$. Moreover, when $\b$ is not too close to zero $n_s$ depends very weakly on $\b$.
For $\b$ close to zero $h$ is allowed to vary from $h\sim15$ up to $h\sim1000$ and farther.
As $\b$ increases the allowed range of $h$ shrinks monotonically and is restricted to $[15,50]$ for $\beta$ close to $1$.

The tensor/scalar ratio $r$ shows a different pattern. For $\b$ close to zero it depends strongly on $h$.
Whereas for values of $\b$ not too close to zero, it depends weakly on both parameters $\b$ and $h$.
Also in this case we observe the monotonic shrinking of the allowed values of $h$ for growing values of~$\b$.

\subsection{Vacuum energy and inflaton mass}

In \cref{fig:EVmI} we show the numerical results obtained from \cref{pm-mass,pm-energy}. We plot the vacuum energy $E_V$ (left) and
the inflaton mass $m_I$ (right) as functions of $\b$ and $h$, again for $N=48,52,56,60$.
The corresponding values of $E_V$ and~$m_I$ are given in terms of the scale of colour shown on the right of every plot.

Because we do not have stringent experimental bounds on $E_V$ and $m_I$, we are interested just in the order of
magnitude of these quantities. We observe that the order of magnitude of $E_V$ depends very weakly on 
$h$ and $N$. Also the dependence on $\b$ is quite weak, as long as we take values of $\b$ not too close to $0$.
Thus, for $\b$ not too close to $0$, the vacuum energy remains about $10^{-4}$ to $10^{-5}$ Planck masses.

On the other hand, the inflaton mass is more sensitive to~$\beta$. Its order
of magnitude is between $10^{-7}$ and $10^{-8}$ Planck masses but for values of $\b$ near to $0$ we have smaller values of $m_I$.

\subsection{Other branch of the potential\label{sect:sb}}

Until now we have considered the slow-roll regime for branch I of the potential, \ie\ $0\le\phi<\infty$.
Let us briefly consider branch II, \ie\ $-\infty<\phi\le0$.
Investigation of this branch is of particular interest because the most interesting cosmological solutions one can
obtain for the exact solvable model with $ h=1/\sqrt{3}$ are defined in branch II of the potential~\cite{Mignemi:2014kea}.

In terms of the parametrization~\eqref{nv}, region II corresponds to $0<Y\le1$.
The slow-roll parameters $\ep$ and $\eta$ are still given by \cref{sr} but now the condition for inflation $\ep\le1$
requires
\be%
\frac{\b-h}{\b(1-\b h)}\le{}Y\le1\0,
\ee%
which can be satisfied only if $h<\b$.
It follows that \mbox{$h=\mathcal{O}(1)$}. One can easily see from \cref{sr,r,ns} that these values of $h$
are not only incompatible with the slow-roll condition $|\eta|\ll1$, but are also completely ruled out by the experimental constraints on~$n_s$.

\afterpage{%
\begin{figure*}[p!]
\centering
\subfloat[$N=48$]{\includegraphics[height=.22\textheight]{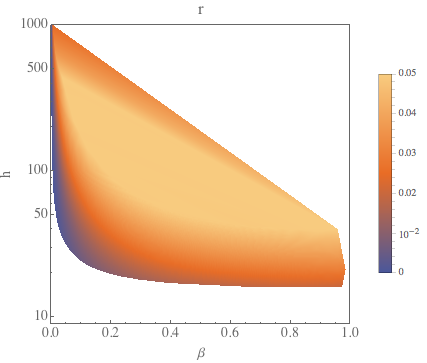}\quad\includegraphics[height=.22\textheight]{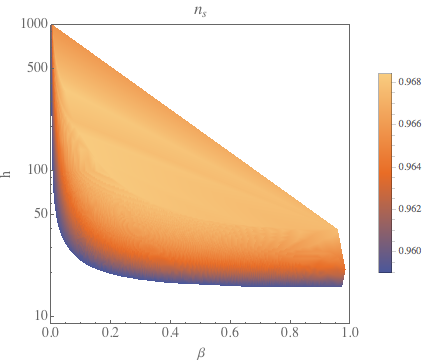}}\\
\subfloat[$N=52$]{\includegraphics[height=.22\textheight]{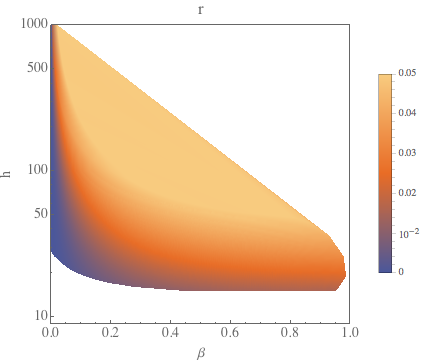}\quad\includegraphics[height=.22\textheight]{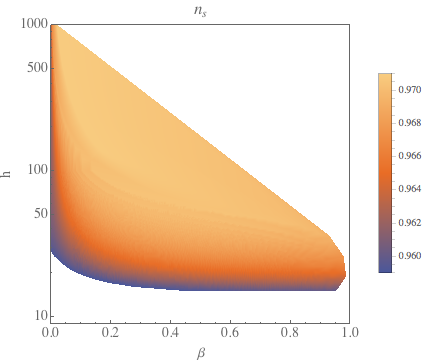}}\\
\subfloat[$N=56$]{\includegraphics[height=.22\textheight]{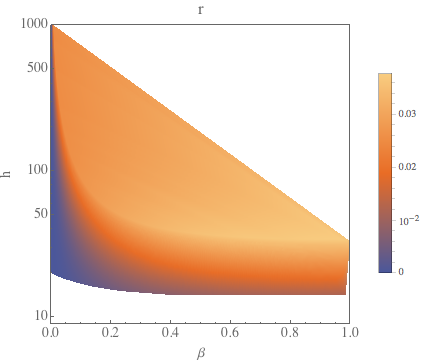}\quad\includegraphics[height=.22\textheight]{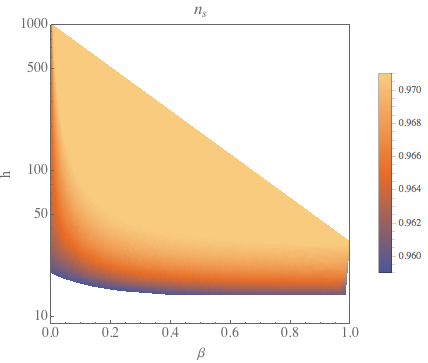}}\\
\subfloat[$N=60$]{\includegraphics[height=.22\textheight]{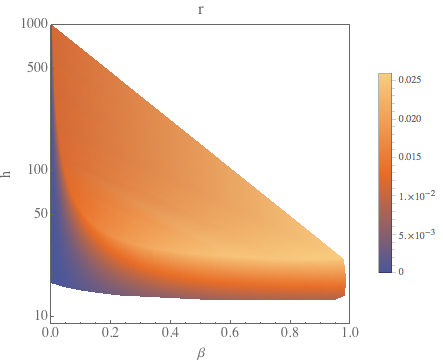}\quad\includegraphics[height=.22\textheight]{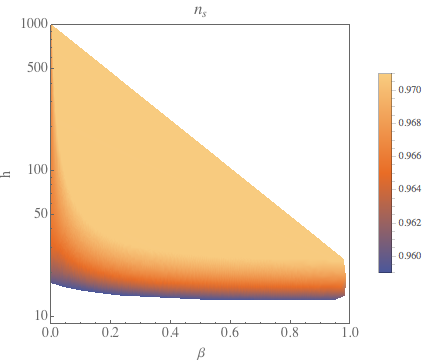}}
\caption[]{Region plots for the tensor/scalar ratio $r$ (left) and the spectral index $n_s$ (right) as functions of the parameters $\b$ and~$h$,
for selected values of the number of $e$-folds $N$. The values of $r$ and $n_s$ are given in terms of the colour scale shown on the right of every plot.}
\label{fig:rns}
\end{figure*}
\begin{figure*}[p!]
\centering
\subfloat[$N=48$]{\includegraphics[height=.22\textheight]{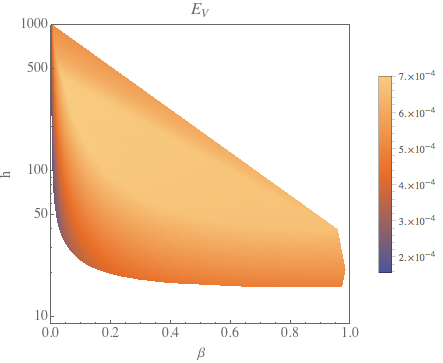}\quad\includegraphics[height=.22\textheight]{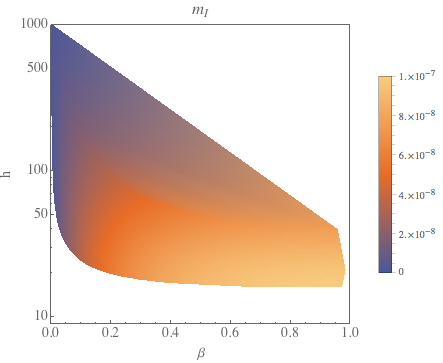}}\\
\subfloat[$N=52$]{\includegraphics[height=.22\textheight]{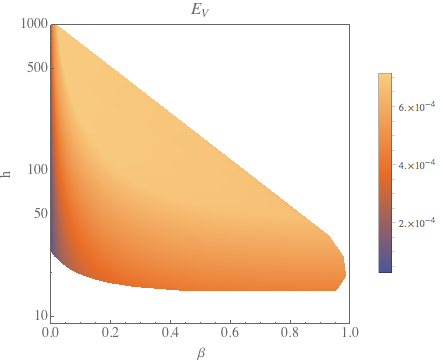}\quad\includegraphics[height=.22\textheight]{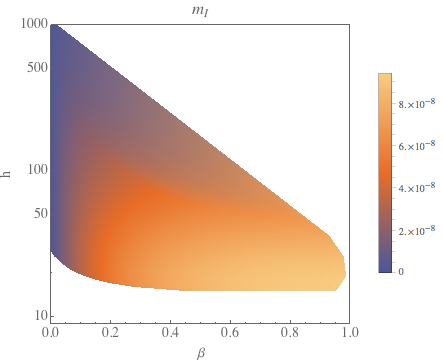}}\\
\subfloat[$N=56$]{\includegraphics[height=.22\textheight]{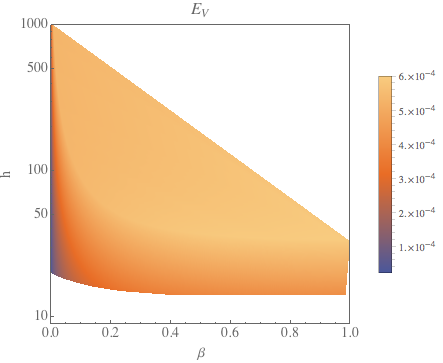}\quad\includegraphics[height=.22\textheight]{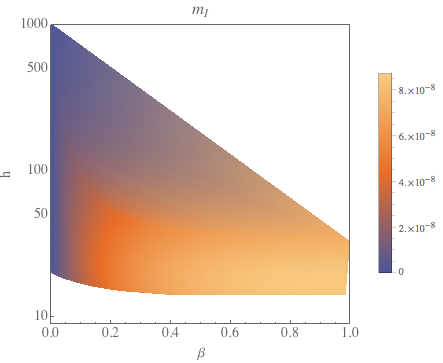}}\\
\subfloat[$N=60$]{\includegraphics[height=.22\textheight]{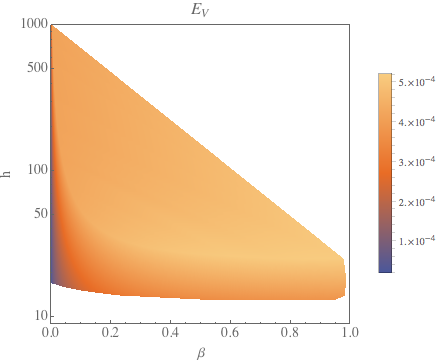}\quad\includegraphics[height=.22\textheight]{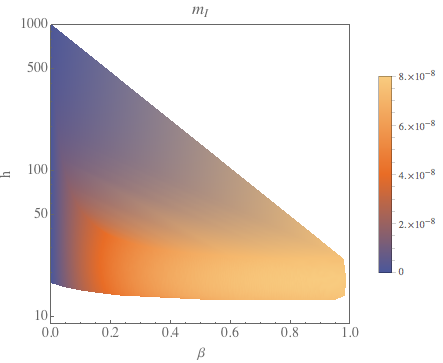}}
\caption[]{Region plots for the vacuum energy $E_V$ (left) and the mass of the inflaton $m_I$ (right), in Planck units, as functions of the
parameters $\b$ and $h$, for the selected values of the number of $e$-folds $N$.
The values of $E_V$ and $m_I$ are given in terms of the colour scale shown on the right of every plot.}
\label{fig:EVmI}
\end{figure*}
\clearpage}

\newpage\section{Conclusion\label{sect:s6}}

In this paper we have constructed the most general Einstein-scalar gravity model in which the potential
is given by the sum of two exponentials and inflation is generated by a scalar field $\phi$ rolling off from the de
Sitter maximum of the potential $V(\phi)$. These models are the cosmological counterparts of holographic models
used to describe hyperscaling violation in the ultraviolet~\cite{Cadoni:2011yj,Cadoni:2012uf}. We have investigated
inflation in the slow-roll approximation. Our model predicts inflation at energy scales of four to five orders
of magnitude below the Planck scale, whereas the inflaton mass, at the onset of inflation, turns out to be seven
to eight orders of magnitude smaller than the Planck mass. We have shown that our model reproduces correctly,
for a wide range of its parameters the most recent experimental data for the power spectrum of primordial perturbations.

The proposed inflationary model belongs to the class of models in which the potential has a dS regime.
This class of models includes the Starobinsky model and, more generally, the cosmological attractor models.
Our model shares with those several features: (1)~the potential is built as a combination of exponentials,
it predicts (2)~an energy scale of inflation four order of magnitude below the Planck mass, (3)~a ``red'' power spectrum
and (4)~a small tensor/scalar amplitude ratio. On the other hand, our model differs from the Starobinsky one in
a crucial aspect: inflation is not generated, as in Starobinsky model, by a scalar field rolling off from an asymptotically
constant potential, but rather from a local maximum of the potential. This property allows us to interpret the inflaton
as a tachyonic excitation of the dS vacuum and to introduce a second scale of energy in the theory, the mass scale $m_I$,
which is 7--8 order of magnitude below the Planck mass. This hierarchy of scales opens the intriguing possibility that the
origin of the inflaton could be explained by the physics at energy scales 7--8 order of magnitude below the Planck mass.

Our model belongs to the general class of hilltop models and shares with the latter the local behaviour near the maximum
of the potential. However, in our model the potential that is constructed has the sum of two exponentials, therefore the global
behaviour of our inflationary model is sensibly different from usual hilltop models constructed using powers of the inflaton. 
In particular, this results in different predictions for the spectral parameters $r$ and $n_s$ in the region of the $e$-folds $N$ accessible to observations.

We close with a brief comment about the reheating phase and the transition from inflation to the radiation/matter dominated era.
During reheating the energy is transferred from the inflaton to matter fields. This means that there must exist a region
in which the kinetic energy of the inflaton dominates over its potential energy, \eg\ a local minimum of the potential.
It is evident from \cref{fig:potV} that the potential \eqref{pot1} does not have such a region and hence it
cannot be used to describe reheating.
Thus, in order to describe reheating our potential must be matched with continuity at the end of
inflation with some other branch of a potential exhibiting a local minimum. This can be done very easily. In
the $Y$-para\-metrization the point $Y_0$ given by \cref{ooo}, at which the universe exits inflation, is always on
the left of the point $Y_*=1$ at which $V$ cuts the horizontal axis, \ie\ we
have $V(Y_0)>V(Y_*)=0$ and $Y_0<Y_*$. Since the slow-roll approximation is badly broken at $V=0$, the matching
with the branch of the potential with the local minimum must be performed at a point $Y_0<Y<Y_*$.

\appendix
\section{\texorpdfstring{$\pmb{\Cosh\phi}$ model}{Appendix A: Cosh(phi) model}\label{sect:a}}

The model \eqref{pot1} is the most general form of the potential one can obtain imposing conditions \eqref{cond} and
assuming that $V$ is built as a combination of two exponentials without an additive constant term.
When such a constant term (which we call $c$) is present, only the first equation in \eqref{cond1} has to be modified and
becomes $c+a_1+a_2>0$, whereas the second and third equations remain unchanged.
A general solution of the ensuing system is given by $a_1= -a_2 (b_2/b_1)$, $a_1,a_2<0$, $b_1>0$, $b_2<0$ and $c>-a_1-a_2$.

A simple example of this class of potentials is given by
\be\label{pot1a}
V (\phi) =\L^2\left(2- \cosh\mu\phi\right).
\ee%
This potential gives a further example of inflation generated by an unstable de Sitter vacuum.%
\footnote{Notice that a similar potential has been investigated in the context of constant-roll inflation,
which reduces to slow-roll inflation when the rate of roll is small --- see \refcite{Motohashi:2014ppa} and references therein.}
\begin{figure*}[!ht]
\centering
\subfloat[]{\includegraphics[width=.4\textwidth]{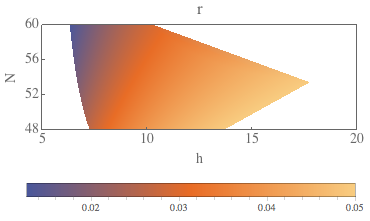}}\quad\subfloat[]{\includegraphics[width=.4\textwidth]{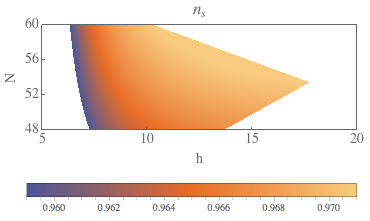}\label{fig:cosha}}\\
\subfloat[]{\includegraphics[width=.4\textwidth]{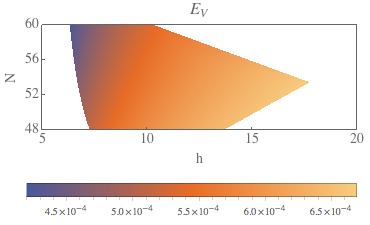}}\quad\subfloat[]{\includegraphics[width=.4\textwidth]{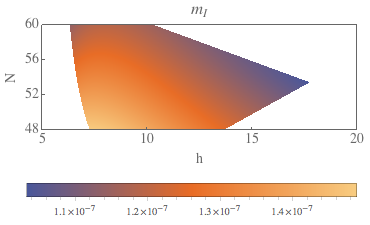}\label{fig:coshb}}
\caption[]{Region plots for (a)~the tensor/scalar ratio $r$, (b)~the spectral index $n_s$, (c)~the vacuum energy $E_V$ and (d)~the mass of the inflaton $m_I$ as functions of the scale parameter $h$ and the number of $e$-folds $N$. $E_V$ and $m_I$ are in Planck masses. The values of $r$, $n_s$, $E_V$ and $m_I$ are given in terms of the colour scale shown below every plot.}
\label{fig:cosh}
\end{figure*}
The potential \eqref{pot1a} has a maximum at $\phi=0$, corresponding to an unstable de
Sitter solution with $V(0)=\L^2$, and a corresponding tachyonic excitation.
For $\mu \phi\gg1$, the potential behaves as a pure exponential.
The vacuum energy and inflaton mass, expressed in terms of $h$ and $\l$, defined as in \cref{pm-mass,pm-energy}, are
\be\label[plural]{pm1}
M^2_I = -\frac{16\pi}{3}\frac{\l^4}{h^2}m_P^2,\quad E_V= \l\,m_P.
\ee%

Introducing the variable $Y= e^{\mu\phi}$, the slow-roll parameters $\ep$ and $\eta$ take the form
\be\label[plural]{sr1}
\ep=\frac{1}{3 h^2}\left(\frac{Y^2-1}{Y^2-4Y+1}\right)^2,
\eta=\frac{2}{3 h^2}\frac{Y^2+1}{Y^2-4Y+1}-\ep.
\ee%
The slow-roll parameter $\ep$ is zero on the maximum of the
potential ($Y=1$). Moreover, we have $0\le\ep\le1$ for \mbox{$1\le Y\le Y_0$}, where
\be\label{ooo1}
Y_0=\frac{2 \sqrt{3}h+ \sqrt{1+9 h^2}}{\sqrt{3}h+1}.
\ee%
For $Y<Y_0$ we have inflation, whereas for
$Y\ge Y_0$ we have $\ep\ge1$ and the universe
exits inflation. One can easily check
that during inflation we always have $1\le Y <2+\sqrt{3}$.
Conversely, the parameter~$\eta$, which gives a measure of the
curvature of the potential, is not small in general, but is of order~$h^{-2}$.

Also for these models the simplest way to satisfy the usual slow-roll
conditions for inflation, $\ep,|\eta|\ll1$, is to choose $h\gtrsim10$,
so that $\eta\approx 10^{-2}$ as well as $\ep\approx10^{-2}$.

The number of $e$-folds $N$ is given by
\be\label{klh1}
\frac{(1+Y)}{(Y(Y-1))^{1/3}}= A\,e^{2N/9h^2},\quad A := \frac{1+Y_{0}}{(Y_{0}(Y_{0}-1))^{1/3}}.
\ee%

In the slow-roll approximation the spectral parameters $P^{1/2}_R$, $r$ and $n_s$ expressed in terms of $N$ are,
\bsube[paramcosh]
&P^{1/2}_R (N) =2 h \l^2\frac{\left(4Y -Y^2-1\right)^{3/2}}{(Y^2-1) Y^{1/2}}\label{pr1a}\\
&r (N) = 16 \ep (N)= \frac{16}{3 h^2}\left(\frac{Y^2-1}{4Y -Y^2-1}\right)^2,\label{r1a}\\
&n_s (N) = 1 -4\ep (N) +2\eta (N)\0\\&\phantom{n_s (N)} = 1- \frac{3}{8}r (N)+\frac{4}{3 h^2}\frac{Y^2+1}{Y^2-4Y+1},\label{ns1a}
\esube%
where $Y=Y(N)$ is defined implicitly as a function of $N$ by \cref{klh1}.

\cref{fig:cosh} shows that there exist values of $h$ for which the model
correctly reproduces the results of observation~\cite{Ade:2015tva} with $N=[48,60]$.
Moreover, it predicts the vacuum energy to be four orders of magnitude below the Planck scale
and the mass of the inflaton seven orders of magnitude smaller than the Planck mass.

\bibliographystyle{apsrev41b}   
\pdfbookmark{\refname}{References}\bibliography{references}

\end{document}